\documentclass[%
 reprint,
superscriptaddress,
 amsmath,amssymb,
 aps,
]{revtex4-2}

\usepackage{graphicx}
\usepackage{dcolumn}
\usepackage{bm}
\usepackage{hyperref}

\newcommand{\myAuthor}[1]{#1}
\usepackage{xcolor}

\usepackage{acronym}
\usepackage{placeins}

\begin{document}

\preprint{Draft}

\title{
Large-scale cooperative sulfur vacancy dynamics in two-dimensional MoS\textsubscript{2} from machine learning interatomic potentials
}

\author{Aaron Flötotto}
\affiliation{%
  Technische Universität Ilmenau, Institute of Physics,\\
  Weimarer Straße 25, 98693 Ilmenau, Germany
}%
\affiliation{
  Institute of Micro- and Nanotechnologies,\\
  Ehrenbergstraße 29, 98693 Ilmenau, Germany
}

\author{Benjamin Spetzler}
\affiliation{
  Christian-Albrechts-Universität zu Kiel, Institute of Materials Science,\\
  Kaiserstraße 2, 24143 Kiel, Germany
}%

\author{Rose von Stackelberg}
\affiliation{%
  Technische Universität Ilmenau, Institute of Physics,\\
  Weimarer Straße 25, 98693 Ilmenau, Germany
}%
\affiliation{
  Institute of Micro- and Nanotechnologies,\\
  Ehrenbergstraße 29, 98693 Ilmenau, Germany
}

\author{Martin Ziegler}
\affiliation{
  Christian-Albrechts-Universität zu Kiel, Institute of Materials Science,\\
  Kaiserstraße 2, 24143 Kiel, Germany
}%

\author{Erich Runge}
\affiliation{%
  Technische Universität Ilmenau, Institute of Physics,\\
  Weimarer Straße 25, 98693 Ilmenau, Germany
}%
\affiliation{
  Institute of Micro- and Nanotechnologies,\\
  Ehrenbergstraße 29, 98693 Ilmenau, Germany
}

\author{Christian Dreßler}
\email{christian.dressler@tu-ilmenau.de}
\affiliation{%
  Technische Universität Ilmenau, Institute of Physics,\\
  Weimarer Straße 25, 98693 Ilmenau, Germany
}%
\affiliation{
  Institute of Micro- and Nanotechnologies,\\
  Ehrenbergstraße 29, 98693 Ilmenau, Germany
}

\date{\today}

\begin{abstract}
The formation of extended sulfur vacancies in MoS\textsubscript{2} monolayers is closely associated with catalytic activity and may also be the basis for its memristive behavior. 
Nanosecond-scale molecular dynamics simulations using machine learning interatomic potentials (MLIPs) reveal key mechanisms of cooperative vacancy transport, including incorporation of vacancies into clusters of arbitrary size. 
The simulations provide a coherent atomistic explanation for irradiation-induced vacancy patterns observed experimentally, especially the formation of line defects spanning tens of nanometers.
Results and performance are compared of two MLIP frameworks: (i) on-the-fly learning with Gaussian approximation potential, and (ii) fine-tuning of an equivariant foundation model. 
\end{abstract}


\maketitle

\acrodef{MD}{molecular dynamics}
\acrodef{DFT}{density functional theory}
\acrodef{NEB}{nudged elastic band}
\acrodef{MLIP}{machine learning interatomic potential}
\acrodefplural{MLIP}[MLIPs]{machine learning interatomic potentials}
\acrodef{RDF}{radial distribution function}
\acrodef{MSD}{mean squared displacement}
\acrodefplural{MSD}[MSDs]{mean squared displacements}
\acrodef{AIMD}{\textit{ab initio} molecular dynamics}
\acrodef{GAP}{Gaussian approximation potential}
\acrodefplural{GAP}[GAPs]{Gaussian approximation potentials}
\acrodef{GPR}{Gaussian process regression}
\acrodef{ACE}{atomic cluster expansion}
\acrodef{GNN}{graph neural network}
\acrodef{SCF}{self-consistent field}
\acrodef{kMC}{kinetic Monte Carlo}
\acrodef{TEM}{transmission electron microscopy}

\acresetall 
\section{Introduction}
Sulfur vacancy migration plays a central role in a wide range of phenomena in two-dimensional (2D) MoS\textsubscript{2}, with relevance to applications including catalysis, chemical sensing, photoluminescence, and electronic transport~\cite{Pang2024, Jin-2024ADFM, Fei-2023ADFM, Chen2024, Hu2021, Li2019}.
In emerging device concepts for next-generation computing, such as memtransistors and memristive elements based on 2D MoS\textsubscript{2}~\cite{Li2018,Feng2021,Fu2023,Jadwiszczak2019,Lee2020,Liu2024,MigliatoMarega2023,Sangwan2015,Sangwan2018,Wang2019,Yang2024,Yuan2021,Zhu2023,Zou2024}, the redistribution of sulfur vacancies within the conduction channel has been widely implicated as a key mechanism underlying their functional behavior~\cite{Li2018,Jadwiszczak2019,Sangwan2015,Spetzler2024,Spetzler2025,Abdel2024,Li2021,Spetzler2022}.

Understanding and ultimately controlling such device functionality requires quantitative insight into the dynamic mechanisms governing vacancy migration and aggregation.
A detailed understanding of these mechanisms must account for the strong dependence of vacancy mobility on the local atomic environment, including variations in defect structure, charge state, and migration pathways~\cite{Li2018,Komsa2013,Chen2018,Gao2021}.
For example, sufficient mobility at room temperature has been shown to require interactions among adjacent vacancies, i.e. vacancy-assisted transport, which can reduce migration barriers from 2-3~eV to 0.6-0.8~eV~\cite{Li2018,Komsa2013,Chen2018,Gao2021}.
However, even with these reduced barrier heights, the associated dynamics still occur on time scales that remain inaccessible to conventional \ac{MD} methods.
Much longer simulation times are particular desirable because the unusual vacancy dynamics in MoS\textsubscript{2} has experimentally been observed to lead to the agglomeration of vacancies into clusters, which preferentially form line defects~\cite{Komsa2013}.
At high vacancy concentrations typical for device applications, or elevated temperatures, vacancy clusters can span up to tens of nanometers~\cite{Chen2018}.
These extended defect structures provide locally conductive pathways for charge transport and further reduce the barrier of vacancy migration~\cite{Komsa2013,Chen2018,Gao2021,Le2014,Enyashin2013,Garcia-Esparza2022,Ryu2016}, making them directly relevant to electronic device functionality. 

In the past, several atomistic simulation methods have been applied to understand sulfur vacancy dynamics in MoS\textsubscript{2}.
\Ac{DFT} has been widely used to calculate electronic properties, formation energies, and migration barriers of selected defect configurations~\cite{Komsa2013,Chen2018,Le2014,Garcia-Esparza2022,Zhou2017,Yang2019,Wang2016,Sensoy2017,Lukashev2024,Liu2013,Komsa2015,Komsa2012,Hong2015,Gusakov2022}.
While offering high accuracy, these methods are computationally expensive, limiting them to static analyses of migration barriers or \textit{ab initio} \ac{MD} simulations on picosecond timescales~\cite{Cohen2012,Jones2015}.
Consequently, they are unable to fully capture the relatively slow dynamic evolution of sulfur vacancy arrangements. 
Purely static analyses of defect dynamics often fail to capture individual low-energy migration pathways, resulting in discrepancies with experimental observations~\cite{Sensoy2017}.
\Ac{kMC} models based on these \ac{DFT} energy barriers extend accessible time and length scales by orders of magnitude~\cite{Wang2024,Wang2022,Wang2024a}, but rely on predefined migration events~\cite{Momeni2020}.
In MoS\textsubscript{2}, the strong dependence of vacancy migration barriers on the local atomic environment~\cite{Wang2022} and the cooperative character of vacancy transport make it extremely challenging to construct accurate models for anything beyond the simplest defect configurations.
Other studies have addressed vacancy dynamics in MoS\textsubscript{2}, using classical \ac{MD}~\cite{Gao2021, Ostadhossein2017, Ponomarev2022, Mohammadtabar2023, Dang2025, Wang2025} and tight-binding MD~\cite{Enyashin2013,Gali2020}.
Classical and tight-binding MD can simulate nanosecond-scale dynamics in large systems.
They have been valuable for exploring cluster formation and local rearrangements but rely on parameterized force fields that may not generalize to unforeseen defect configurations~\cite{Momeni2020}.
Although these classical force fields have been shown to achieve good accuracy for migration barriers around very small sulfur vacancy clusters, they can fail to predict energy barriers for atomic jumps within larger vacancy clusters at \ac{DFT} accuracy~\cite{Gao2021}.
For instance, a previous study addressing similar issues as the present work employed a classical reactive force field and successfully demonstrated the formation of triple-line vacancies through the combination of a double vacancy and an additional vacancy~\cite{Gao2021}.
However, it failed to account for the incorporation of additional vacancies required to form extended line defects. This limitation stems from the fact that the activation energies for low-barrier vacancy jumps were consistent with DFT calculations only in the case of double vacancies.
Despite their respective strengths and the insights \ac{kMC} and classical MD approaches have provided, capturing the structural and dynamic evolution of extended defect structures remains challenging. 

We address this challenge by performing \ac{MD} simulations of MoS\textsubscript{2} using specifically trained \acp{MLIP}, an emerging approach for modeling large-scale vacancy migration and defect evolution.
\Acp{MLIP} have recently demonstrated near-DFT accuracy across nanosecond timescales in similar systems, offering a viable compromise between accuracy and computational efficiency~\cite{Han2023,Liu2024a,Siddiqui2024,Unke2021}.
Among the available approaches, the \ac{GAP}~\cite{Bartok2010} and the MACE \ac{GNN}~\cite{Batatia2022} exhibit complementary strengths: GAP enables uncertainty-based on-the-fly training~\cite{Jinnouchi2019,Jinnouchi2019a}, while MACE supports transfer learning from general-purpose foundation models~\cite{Batatia2023, Deng2025, Angeletti2025}.
In this work, we compare these methods by training GAP models during \ac{MD} simulations of MoS\textsubscript{2} and fine-tuning MACE models to MoS\textsubscript{2} \ac{DFT} data starting from the MACE MP-0 foundation model~\cite{Batatia2023}.
For both approaches, we evaluate different training data selection strategies and apply the resulting \acp{MLIP} to simulate sulfur vacancy migration over multiple nanoseconds, capturing structural evolution processes that remain inaccessible to conventional atomistic techniques.
Our simulations reveal the underlying mechanism of sulfur vacancy dynamics in MoS\textsubscript{2} monolayers and provide an atomistic explanation for the experimentally observed aggregation of small defect clusters into extended line defects spanning tens of nanometers.


\section{Results and Discussion}
\subsection{Training of \ac{MLIP} models}
\label{sub:results_mliptraining}
First of all, we outline the process of generating \acp{MLIP} that can later be applied to perform \ac{MD} simulations of defective supercells on the time scale of nanoseconds with close to \ac{DFT} accuracy.
As described in detail in the Method Section~\ref{sub:methods_mlff}, we employed two different machine learning architectures to construct \acp{MLIP}.
The first is the equivariant \ac{GNN} MACE~\cite{Batatia2022} using the atomic cluster expansion to describe structural environments \cite{Drautz2019}.
Secondly, we applied \acp{GAP}~\cite{Bartok2010} using an on-the-fly training approach and atomic descriptors as implemented in VASP \cite{Jinnouchi2019, Jinnouchi2019a, Jinnouchi2020}.
For each \ac{MLIP} architecture, two different approaches for generating training data were tested.
All models are fine-tuned to \ac{DFT} forces and energies, but differ in the number of structures in the training set and how these structures were obtained.
The test errors of selected \acp{MLIP} models are reported in detail in Section~\ref{sub:methods_testerror}.
With one exception the resulting force and energy test errors of the various \acp{MLIP} models are within the expected range for these \ac{MLIP} architectures: a few meV for the energies and a few 10~meV~\AA$^{-1}$ for the forces~\cite{Batatia2022, Jinnouchi2019}.
Only a variant of the \ac{GAP} model that has been fine-tuned on supercells with four quite different defects shows comparably high test errors for forces.
We interpret this as trade-off between transferability and accuracy.

\subsection{Evalution of \ac{MLIP} predicted potential energy curves for cooperative, vacancy-assisted hopping}
\begin{figure*}
\centering
\includegraphics[width=0.9\textwidth]{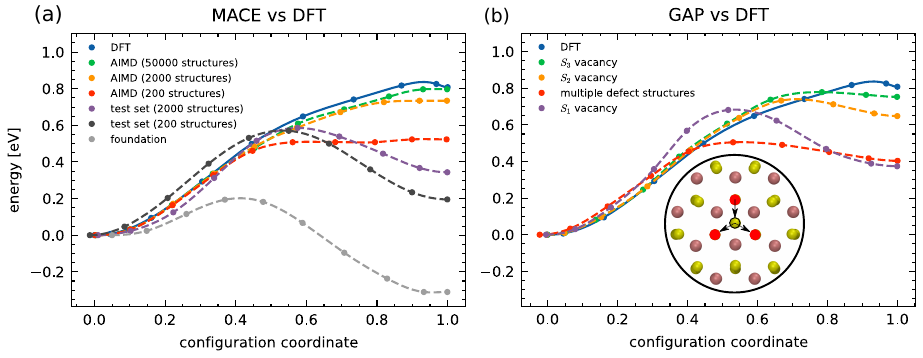}   
\caption{
Potential energy curves for the defect jump shown in the inset of (b)
calculated using the \ac{NEB} method.
A comparison between results calculated with \ac{DFT} and MACE or \ac{GAP}
\acp{MLIP} trained on different datasets (see
Section~\ref{sub:methods_mlff}) is
shown in (a) and (b), respectively.
MACE \acp{MLIP} were trained on data from \ac{AIMD} simulations (label: ``AIMD'') or from \ac{DFT} calculations of structures from an \ac{MLIP} \ac{MD} simulation using the MACE MP-0 foundation model (label: ``test set'').
The \ac{GAP} models differ in the defect structure present during the on-the-fly training \ac{MD} simulations (see Section~\ref{sub:methods_mlff} for more details).
In addition to the \ac{NEB} data points, we show spline interpolations
through them as a guide for the eye.
In the inset, sulfur atoms are displayed in yellow and molybdenum atoms in purple.
The moving sulfur atom is indicated with a black ring and the surrounding
vacancies are indicated with red circles.
The potential energy curves shown here correspond to the jump indicated
by the vertical arrow.
The non-vertical arrows represent possible additional components of jumps
that are observed in \ac{MD} simulations and that share the same potential
energy curve as the first component.
The configuration coordinate is defined in Equation~(\ref{eq:confco}) in 
Section~\ref{sub:methods_neb}.
}
\label{fig:neb}
\end{figure*}

The subject of this work, the large-scale dynamics, is determined by rather rare events.
In contrast, the test errors reported in Table~\ref{tab:testtrainingerrors} are based on equidistantly selected snapshots from \ac{MLIP} \ac{MD} simulations, which were reevaluated using \ac{DFT}.
As a result, rare events---such as vacancy jumps---are unlikely to be included in the test set.
Consequently, the test errors primarily reflect the accuracy of the \acp{MLIP} on low-energy structures that frequently occur during \ac{MD} simulations in the form of vibrations of atoms around their equilibrium positions and deformations of the local environment.
Therefore and in order to assess the accuracy of our \acp{MLIP} in a more targeted way, we calculate and compare potential energy curves of a particular sulfur defect jump within the \ac{NEB} framework, a standard approach --- usually of \ac{DFT} --- to find the transition state and the lowest energy barrier between two prescribed configurations \cite{Jonsson1998}.
Specifically, we look at the jump of a sulfur atom to an interstitial site neighboring a pair of two adjacent vacancies in one of the sulfur layers.
This jump has previously been studied using high-resolution \ac{TEM} and \ac{DFT}~\cite{Komsa2013} and is illustrated by the vertical solid arrow in the inset of \textbf{Figure~\ref{fig:neb}} (b).
After the jump, the sulfur atom is positioned at an interstitial site centrally between three sulfur vacancies, with two of these vacancies present prior to the atom's jump and the third vacancy being at the site from which the atom originated.
In our \ac{MD} simulations and actual experiments, this jump is typically followed by another jump of the same sulfur atom from the interstitial site back to the original vacancy or to one of the previously existing vacancies as indicated by the non-vertical arrows.
These subsequent jumps are characterized by the same potential energy curve as the first jump to the interstitial site due to the symmetry of the structure.
We note in passing that the result of the two equivalent jumps can be considered as vacancy-assisted hopping~\cite{Ritort2003, Jaeckle1999}: a sulfur atom and a vacancy switch positions in the presence of another 'assisting' vacancy.  

In the following, we compare the NEB potential energy curves for various different \acp{MLIP} versions to reference \ac{DFT} calculations for this jump (full blue line in Figure~\ref{fig:neb} (a) and (b)).
The DFT energy difference between the initial and final configurations is 0.81~eV. 
The energy barrier for the jump to the interstitial site is 0.83~eV, which makes the latter weakly meta-stable with a very small energy barrier of about 0.02~eV for the backward jump from the interstitial site back to the original lattice site or one of the other two equivalent lattice sites.

The results for MACE and \ac{GAP} \acp{MLIP} are shown in Figure~\ref{fig:neb} (a) and (b), respectively.
Most noticeably, the MACE foundation model (gray symbols) yields even qualitatively wrong results: It predicts the energy of the interstitial sulfur defect to be considerably lower than that of the initial configuration with only two vacancies.
All fine-tuned MACE models perform significantly better than the foundation model.
However, the agreement with the \ac{DFT} \ac{NEB} calculations depends quite strongly on the choice of training set.
Black and purple symbols mark results for  MACE \acp{MLIP} fine-tuned to test sets of 200 and 2,000 \ac{DFT} calculations for configurations obtained as snapshots from a \ac{MD} trajectory calculated with the foundation model.
While these models accurately reproduce the potential energy curve near its minimum, they significantly deviate from the \ac{DFT} reference values for energy barriers (0.57~eV and 0.59~eV, respectively, for the forward jump and 0.38~eV and 0.24~eV, respectively, for the backward jump from the interstitial site into one of the vacancies).
The poor description of the transition region is somewhat surprising, because as the ground state of the foundation model, it should actually be overrepresented in the training sets.
A possible explanation lies in our observation that the MACE foundation \ac{MD} trajectory includes numerous other unphysical transition states. 
For example, it predicts rather frequent sulfur diffusion through the molybdenum layer from one side of the MoS\textsubscript{2} sheet to the other.

Qualitatively almost correct and quantitatively much better results are obtained by fine-tuning the MACE model on 200, 2000, and 50,000 snapshots obtained from \ac{AIMD} simulations (red, orange, and green symbols, respectively).
For a completely different material system, this approach has been shown to yield highly accurate \acp{MLIP} over the whole configurational space using a relatively small fine-tuning dataset of around 1~ps of \ac{AIMD} data~\cite{Grunert2025}.
In the case of MoS\textsubscript{2}, we find that the accuracy at regions farther away from the potential energy minimum depends sensitively on the size of the fine-tuning set.
Specifically, the \ac{MLIP} fine-tuned on 200 \ac{AIMD} snapshots predicts an energy difference of 0.52~eV  between the initial and final configuration, while the one fine-tuned on 2,000 snapshots yields a value of 0.73~eV.
With a substantially larger fine-tuning set size of 50,000 snapshots using every second snapshot of a 100~ps long \ac{AIMD} trajectory with time step of 1~fs, the MACE model closely approaches the DFT value (0.81~eV) for this energy difference with 0.80~eV and generally predicts the potential energy curves very accurately.
None of the these three AIMD-fine-tuned MACE models predict the final configuration of the jump to be metastable. 
Given that \ac{DFT} predicts a backward energy barrier of only $0.03 \ \mathrm{eV} = k_B \ 348 \ \mathrm{K} $, this will cause only very minor errors in all our \ac{MD} simulations, which were performed for much larger temperatures.

To summarize the results for MACE \acp{MLIP} shown in Figure~\ref{fig:neb}~(a): 
(\textit{i}) Using a large training set sampled from \ac{AIMD} simulations, we are able to fine-tune a model that very accurately describes these sulfur jumps.
(\textit{ii}) The MACE foundation model and all models fine-tuned on structures taken from \ac{MD} simulations using the foundation model strongly underestimate the potential energy of the transition state.
This underprediction of migration barriers has recently been reported for a variety of chemical systems and multiple universal \acp{MLIP} including the MACE MP-0 foundation model by \myAuthor{Deng} \textit{et al.}~\cite{Deng2025}.
Our observations for MoS\textsubscript{2} fully confirm their emphasis on the importance of sampling non-equilibrium structures during the  training. 
However, \myAuthor{Deng} \textit{et al.}~\cite{Deng2025} did not see, or at least did not report the difference between sampling structures for the fine-tuning dataset from \ac{MD} simulations using the foundational \ac{MLIP} or more accurate but slower and more expensive \ac{AIMD} simulations that we observed.

Next, we turn towards the performance of the \ac{GAP} \acp{MLIP}, see Figure~\ref{fig:neb}~(b).
As described in Section~\ref{sub:methods_mlff}, these models were trained using an on-the-fly learning approach during \ac{MD} simulations of different defect configurations.
The models trained starting with a $S_2$ vacancy pair (orange symbols) or a $S_3$ vacancy triplet (green) neatly  match the \ac{DFT} potential energy curve (blue) near the energy minimum and show only minor deviation near the maximum:
Both predict the interstitial position to be meta-stable. The energy barriers of 0.74~eV and 0.78~eV for the forward jump are close to the DFT value (0.83~eV), but slightly less accurate than the best MACE model (0.80~eV).
The better performance of the vacancy-triplet-trained \ac{GAP} \acp{MLIP} may be due to the fact that its on-the-fly learning involved more jumps.
We checked that the jumps occurring in the on-the-fly training runs for these \acp{MLIP} involve the same process as studied in the \ac{NEB} calculations.
In contrast, for the model trained starting with only a single vacancy in the structure ($S_1$, purple) no jumps at all occurred during the training.
Not surprisingly, it shows much larger deviations from the \ac{DFT} potential energy curve than the previous two.
Also the force field trained on multiple, structurally different vacancy clusters (red symbols) is accurate only near the energy minimum and, thus, not suitable for accurately describing the defect dynamics in \ac{MD} simulations.

\subsection{Discussion of other hopping processes\label{subsec-otherhopps}}
\begin{figure}[htpb]
    \centering
    \includegraphics[width=.4\textwidth]{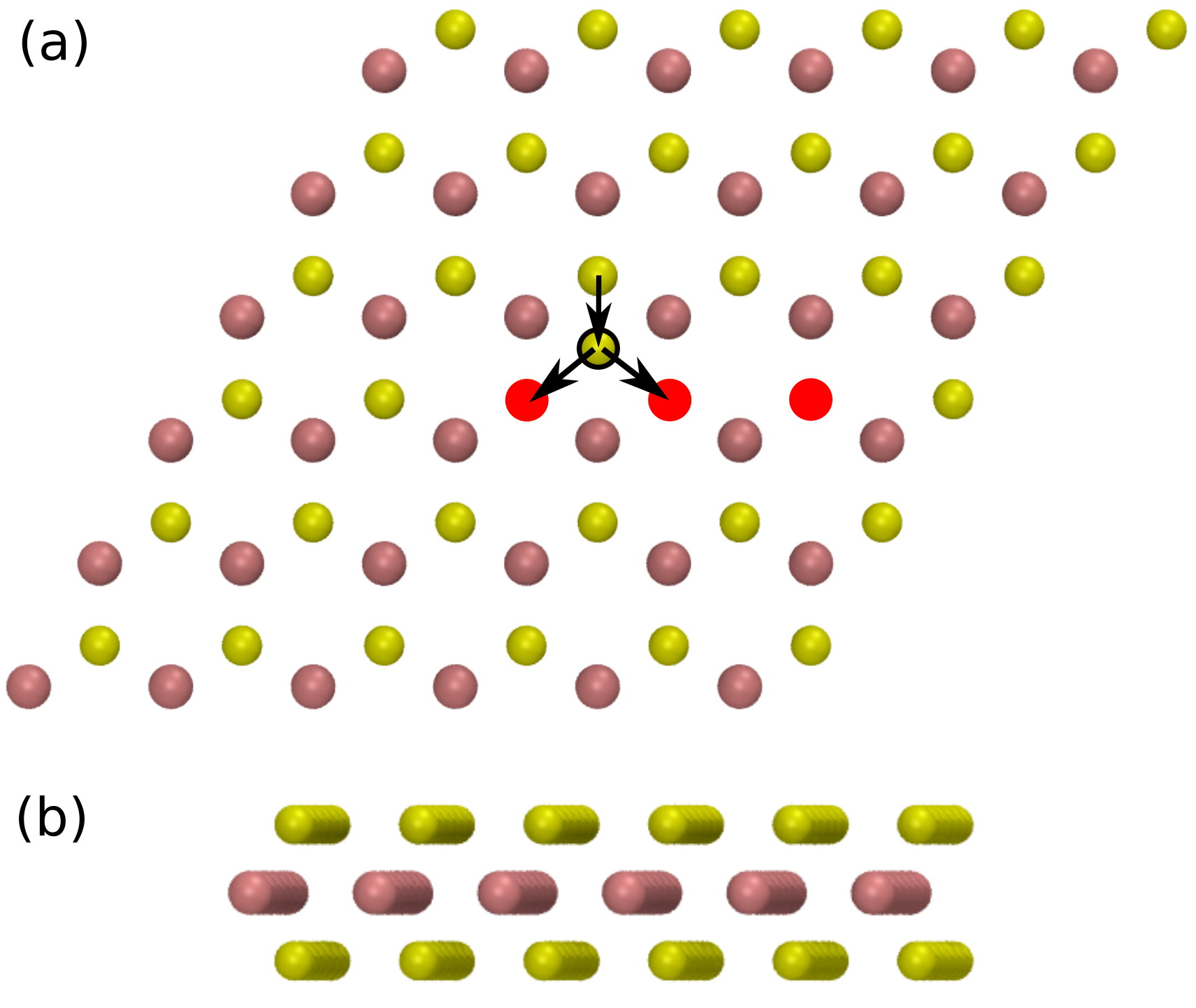}
    \caption{Top (a) and side (b) view of a MoS\textsubscript{2} monolayer in the hexagonal 2H phase. Sulfur and molybdenum are displayed in yellow and purple, respectively.
    The annotations in (a) mark the final jump discussed in Section~\ref{subsec-otherhopps}: A sulfur atom hopping into a line of three vacancies (red circle: vacancy, black ring: jumping atom on the intermediate, interstitial site).
    }
    \label{fig:structure}
\end{figure}

\myAuthor{Wang} \textit{et al.}~\cite{Wang2022} conducted a comprehensive analysis of DFT-calculated energy barriers for sulfur jumps into a vacancy for the use in \ac{kMC} calculations. 
They considered all 256 possible configurations of the eight neighboring sites surrounding both the initial and final positions of the jumping sulfur atom.
\myAuthor{Wang} \textit{et al.} identified the jump used in our evaluation in Figure~\ref{fig:neb} as having one of the lowest energy barriers, alongside with related configurations where additional neighboring sites are vacant.
The relatively low barrier is attributed to the presence ('assistance') of a vacancy adjacent to the destination site, which allows the sulfur atom to access the interstitial site.
The latter is energetically favorably only if it is not directly above a molybdenum atom in the layer below, which is the case for every second of these interstitial sites as illustrated in \textbf{Figure~\ref{fig:structure}}.

Two defect jumps similar to the one shown in Figure~\ref{fig:neb} were previously studied by \myAuthor{Ostadhossein} \textit{et al.}\cite{Ostadhossein2017}.
These jumps also originate from a double vacancy configuration, like the one considered in our work.
However, in contrast to the jump analyzed here, the mobile sulfur atom in their study starts from a site farther from the center of the double vacancy, directly adjacent to only one of the two vacancies.
The \ac{DFT}-predicted energy barrier for the jump investigated in our work is lower than those of these alternative jumps, which were reported to have barriers of 1.35~eV and 2.32~eV, depending on the sulfur atom involved~\cite{Ostadhossein2017}.
In earlier studies, ReaxFF reactive force fields have been developed for MoS\textsubscript{2}~\cite{Ostadhossein2017, Ponomarev2022, Mohammadtabar2023, Dang2025, Wang2025}.
\myAuthor{Ostadhossein} \textit{et al.}~\cite{Ostadhossein2017} evaluated the accuracy of their force field by comparing the predicted potential-energy profiles of these jumps to DFT results.
Although the types of the jumps differ, we note that the energy barrier errors predicted by ReaxFF are larger than those from our best \acp{MLIP}, when benchmarked against DFT.

We also calculated the potential energy curve for the jump of a sulfur atom into a single sulfur vacancy.
The minimum energy path for this jump into a single vacancy does not run through the center of the triangle spanned by three sulfur lattice sites, cf.~inset of Figure~\ref{fig:neb}~(b).
Instead, it keeps more distance to the now occupied lattice site and thus runs closer to the neighboring molybdenum site.
Compared to the case of two adjacent vacancies discussed earlier, the energy barrier for this jump is much larger: Our most accurate MACE model yields a value of 1.81~eV.
Thus, according to the exponential Arrhenius law and considering the 1~eV barrier-height difference, the jump of a sulfur atom into one of two adjacent vacancies is far more likely than into a single vacancy at any technologically relevant temperature.

Finally, we calculated using our best MACE \ac{MLIP} model potential energy curves for the jump of one of the sulfur atoms into to a straight line of three vacancies, see Figure~\ref{fig:structure}~(b).
The sulfur atom can jump either into the center of the vacancy line or into one of the outer vacancies.
We find an energy barrier of about 0.8~eV just as the jump into two neighboring vacancies shown in Figure~\ref{fig:neb} for both of these alternatives.
However, the potential energies of the two possible final configurations differ significantly:
The energy of the final configuration with the sulfur atom in the center of the row is 0.28~eV higher than that of the configuration with a sulfur atom occupying one of the outer vacancies.
Consequently, the energy barrier for the reverse jump is 0.28~eV lower, i.e. approximately 0.52~eV.
Energy barriers for jumps within a three-vacancy cluster have previously been calculated using \ac{DFT} and a ReaxFF force field~\cite{Gao2021}.
The values predicted by our \ac{MLIP} ($\sim 0.8$~eV) are in good agreement with those \ac{DFT} results.
In contrast, the ReaxFF force field drastically underestimates the barrier for the sulfur jump into the central vacancy (0.30~eV).
Simultaneously, it overestimates the barrier for sulfur jumps into one of the outer vacancies, predicting 1.37~eV compared to 0.85~eV from \ac{DFT}~\cite{Gao2021}.
Notably, \myAuthor{Gao} \textit{et al.} show that for only two adjacent vacancies, their ReaxFF model yields energy barriers in good agreement with \ac{DFT}.
As a result, their classical \ac{MD} simulations provide valuable insights into the evolution of double vacancies.
However, as the cluster size increases, the ReaxFF force field fails to accurately capture the correct dynamics due to its inaccurate barrier predictions, as will be discussed below in more detail.

\subsection{Evaluation of \ac{MLIP}-predicted radial distribution functions}
\begin{figure}[htpb]
    \centering
    \includegraphics[width=.45\textwidth]{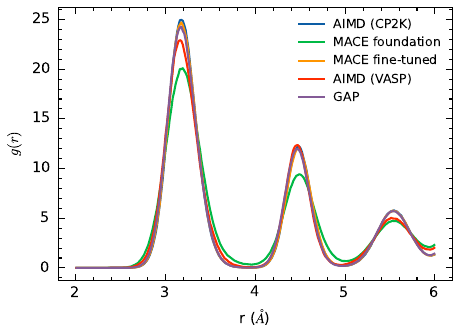}
    \caption{
    Time-averaged S-S radial distribution function $g(r)$ calculated from \ac{MD} simulations of a MoS\textsubscript{2} monolayer with three sulfur vacancies in a row using both, \ac{AIMD} and \acp{MLIP}.
    The \ac{GAP} and MACE fine-tuned \acp{MLIP} examined here are those with the best agreement with DFT in terms of potential energy curves in Figure~\ref{fig:neb}~(a) and (b).
    The \acp{RDF} are calculated as described in Sections~\ref{sub:methods_rdf}.
    }
    \label{fig:rdf}
\end{figure}

\Ac{RDF} of sulfur $g_{\textit{S-S}}(r)$ in a single layer of MoS\textsubscript{2} provide another way to assess the quality of the \acp{MLIP} for \ac{MD} simulations.
Results using both, \ac{DFT} and \acp{MLIP} are shown in \textbf{Figure~\ref{fig:rdf}}.
The first peak at about 3.2~\AA \ corresponds to both, the distance between neighboring sulfur atoms within one of the two sulfur layers visible in Figure~\ref{fig:structure}~(b) and the distance between two sulfur atoms from the two different layers that are on top of each other in Figure~\ref{fig:structure}~(a).
The \acp{RDF} show good agreement between the two \ac{AIMD} simulations performed using CP2K and VASP (see Section~\ref{sub:methods_aimd} for details), which serve as the reference, and the \ac{GAP} and MACE \acp{MLIP}, which produced the most accurate potential energy curves.
Only the MACE foundation model is unable to predict the \ac{RDF} accurately.
The AIMD trajectory generated using CP2K was employed as training data to fine-tune the MACE model, while the VASP \ac{DFT} implementation was used in the on-the-fly learning of the \ac{GAP} model.
Differences observed in the resulting \acp{RDF} from these two AIMD simulations arise from the distinct \ac{DFT} implementations, as described in Section~\ref{sub:methods_aimd}.
However, the difference in peak area between the \ac{AIMD} \acp{RDF} is significantly smaller than the difference with the MACE foundation model, which also fails to produce the correct potential energy surface.

\subsection{Sulfur vacancy dynamics}
After generating \acp{MLIP}, capable of predicting potential energy curves and \acp{RDF} with satisfactory accuracy, we applied them to determine the \ac{MSD} and defect jump rates of various defect structures in monolayer MoS\textsubscript{2} on timescales not accessible with \ac{AIMD}.
The much extended accessible simulation times also allow for the identification of cooperative effects, which are discussed in the next section.   

\begin{figure*}
  \centering
  \includegraphics[width=0.9\textwidth]{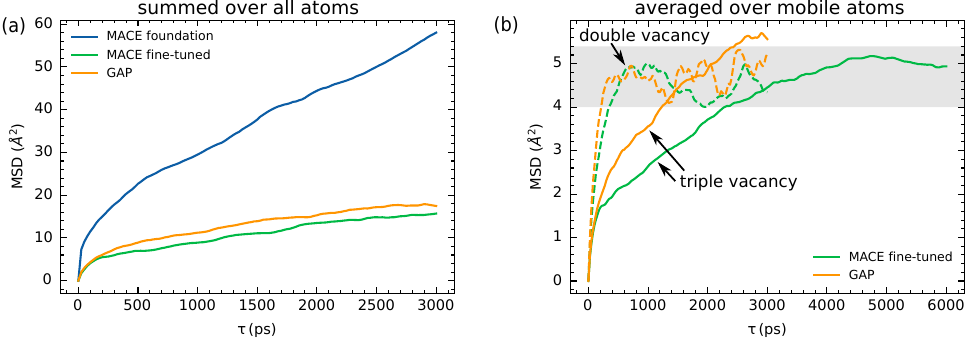}
  \caption{
  (a) Mean squared displacement (MSD) summed over all sulfur atoms calculated as described in Section~\ref{sub:methods_msd} from \ac{MD} simulations of a MoS\textsubscript{2} monolayer starting with three sulfur vacancies in a row using three different \acp{MLIP}.
  The \ac{GAP} and MACE fine-tuned \acp{MLIP} applied here are those with the best agreement with DFT in terms of potential energy curves in Figure~\ref{fig:neb}.
  For reference, we also show results of \ac{MD} simulations using the MACE foundation model, which shows poor agreement with DFT on the potential energy curve in Figure~\ref{fig:neb}~(a) and is thus expected to dramatically overestimate the mobility of sulfur vacancies.
  (b)~\Ac{MSD} of mobile sulfur atoms calculated from \ac{MD} trajectories with two (dashed lines) or three (solid lines) adjacent sulfur vacancies in the initial structure using the same MACE fine-tuned (green) and \ac{GAP} (orange) \acp{MLIP} as in (a).
  The MACE calculation with three vacancies was performed over a significantly longer simulation time (about 35~ns) to demonstrate that it eventually reaches a similar plateau as the simulations with two vacancies (10~ns simulation time).
  }
  \label{fig:msd}
\end{figure*}

\textbf{Figure~\ref{fig:msd}}~(a) shows the sulfur \ac{MSD} of a structure containing three sulfur vacancies computed from \ac{MD} simulations for $T = 1000$~K calculated with the MACE foundation model and the best fine-tuned MACE (green line) and \ac{GAP} (orange) models shown in Figure~\ref{fig:neb}.
As described in Section~\ref{sub:methods_msd}, these \acp{MSD} are corrected for the displacement due to atomic vibrations and thus show only the displacements of mobile sulfur atoms.
For purely diffusive motion, all the conventional MSD and its variants (summed of atoms, corrected for vibrations) would grow linearly with time.

The two MoS\textsubscript{2}-specific \acp{MLIP} show reasonably good agreement with each other.
Both predict the mobile sulfur atoms to move by about 3~\AA\ on average over a time span of 1~ns.
Based on the simulation data, only 3 atoms were observed to migrate between distinct lattice sites.
In contrast, the MACE foundation model (blue line in Figure~\ref{fig:msd}~(a)) predicts much larger displacements in general, more mobile atoms, and a more or less linear increase in the \ac{MSD}, which would suggest unrestricted diffusion of sulfur atoms on this time scale.
Again, we attribute this failure to the systematic softening of the foundation model~\cite{Deng2025}, which leads to an incorrect reduction not only of the energy barrier shown in Figure~\ref{fig:neb} but also of other jump barriers for sulfur atoms.
In particular, in this MACE foundation \ac{MD} simulation, we observe several jumps of sulfur atoms from the upper layer through the molybdenum layer to the bottom layer and vice versa that do not occur in \ac{MD} simulations based on \ac{DFT} or MoS\textsubscript{2}-specific \acp{MLIP} on this timescale.
As shown earlier, the MACE foundation model also predicts a wrong ground state configuration for the sulfur atoms close to a double sulfur vacancy.
However at the simulation temperature of 1000~K, sulfur atoms are not trapped in this configuration due to the relatively low energy barrier of the backwards jump or equivalent  jumps---$\sim 0.5$~eV according to the MACE foundation model as shown Figure~\ref{fig:neb}.
This leads to the huge \ac{MSD} values shown in Figure~\ref{fig:msd}~(a) which continue to grow even after $\tau = 3$~ns.

Figure~\ref{fig:msd}(b) shows the averaged \acp{MSD} of mobile sulfur atoms obtained from MACE fine-tuned and \ac{GAP} \ac{MD} simulations, starting from structures with two or three adjacent vacancies.
When only a single vacancy is present, both \acp{MLIP} predict that sulfur atoms do not undergo lattice jumps, resulting in negligible mobility.
In this case the only displacement of sulfur atom results from their thermal vibration about the equilibrium position yielding \acp{MSD} of roughly 0.1~\AA$^2$.
In contrast, when two or three vacancies are present, significantly larger displacements of sulfur atoms are observed, corresponding to one and three mobile atoms, respectively.

\begin{figure*}[htpb]
    \centering
    \includegraphics[width=0.95\textwidth]{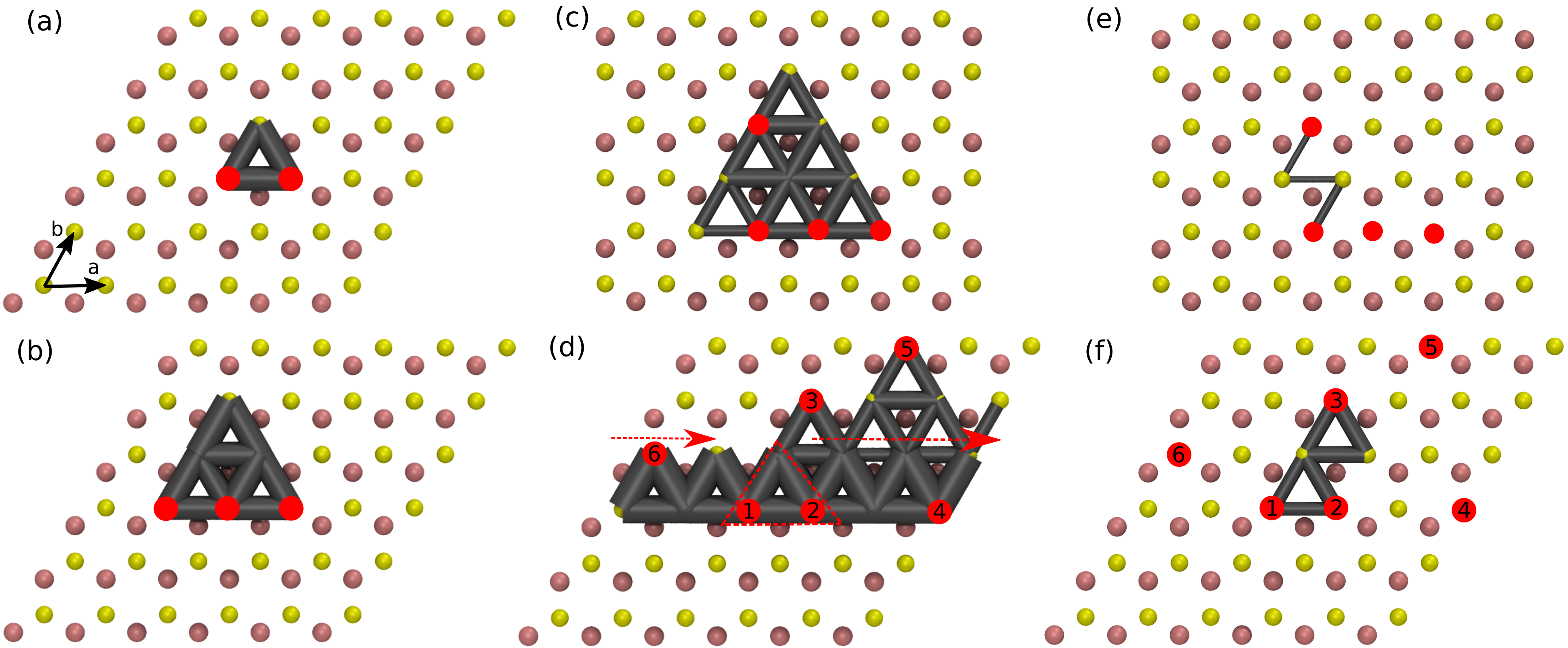}
    \caption{
    Visualization of sulfur atom jumps observed in \ac{MD} simulations with $T = 1000$~K using our most accurate MACE \ac{MLIP}.
    The initial positions of sulfur vacancies are indicated by red circles.
    Jumps between two sites are indicated by gray cylinders, with cylinder radii scaling logarithmically with the number of jumps between the two sites during the \ac{MD} simulation.
    Panels (a) to (d) result from 10~ns long simulations.
    To highlight the importance of long simulation times, panels (e) and (f) show jump rates from only the first 100~ps 
    -- a typical timescale accessible to \ac{AIMD} simulations --
    of the simulations fully shown in (c) and (d).
    }
    \label{fig:jumpvis_mace}
\end{figure*}

For both cases, the \acp{MSD} of mobile atoms reach a plateau at approximately 5~\AA$^2$ (see Figure~\ref{fig:msd}~(b)).
This indicates that the motion of sulfur atoms is spatially constrained in this finite system at the ns timescale.
Notably, the system with two vacancies exhibits a faster convergence of the \ac{MSD}, reaching its plateau after around 0.5~ns.
In contrast, the system with three vacancies continues to show an increasing \ac{MSD} for several nanoseconds before leveling off.
Thus, in the system with two vacancies, the single mobile sulfur atom appears to be confined to a smaller region, leading to a more rapid convergence of the \ac{MSD}.
We examine this behavior in greater detail later, in the discussion of \textbf{Figure~\ref{fig:jumpvis_mace}}.
We emphasize that these studies can neither be done within the computationally too expensive \ac{AIMD} framework nor with the inaccurate MACE foundation model without fine-tuning. 

\begin{figure}[htpb]
    \centering
    \includegraphics[width=.45\textwidth]{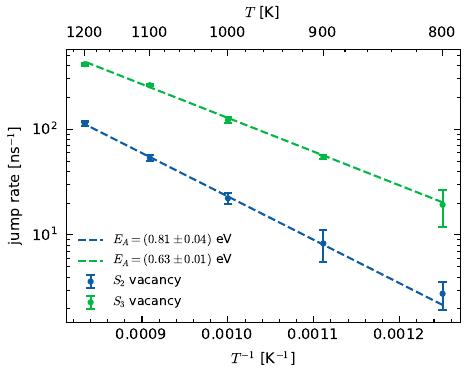}
    \caption{
    Arrhenius plot for the jump rate in 20~ns long \ac{MD} simulations of MoS\textsubscript{2} sheets with two or three adjacent sulfur vacancies as shown in Figure~\ref{fig:jumpvis_mace} at different temperatures.
    The simulations were performed with our most accurate MACE \ac{MLIP} that was trained on \ac{AIMD} data of defective MoS\textsubscript{2}.
    The uncertainties of the data points are roughly estimated by dividing the \ac{MD} trajectories into 4 parts each and calculating the mean of the jump rate in each part and its standard error.
    }
    \label{fig:jumps_arrhenius}
\end{figure}

The capability to run highly accurate nanosecond-scale \ac{MD} simulations now allows us to analyze the statistics of the relatively rare sulfur atom jumps.
We performed 20~ns long simulations at temperatures of 800, 900, 1000, 1100, and 1200~K using our best fine-tuned MACE model
with initial structures containing two or three vacancies arranged in a row.
\textbf{Figure~\ref{fig:jumps_arrhenius}} shows the number of sulfur jumps per nanosecond, i.e.\ the jump rate, and error estimates as function of temperature.
The number of jumps was determined by projecting each atom to its nearest lattice site and counting the changes of the nearest lattice site between snapshots that were sampled at 50~fs intervals; the statistical error was estimated by dividing the 20~ns long \ac{MD} trajectories into 4 parts and calculating the jump rate for each part separately.
Note that only about 50 jumps occurred during the 20~ns long double-vacancy simulation at 800~K.

In accordance with the Arrhenius law, the logarithm of the jump rate depends linearly on the inverse temperature allowing us to estimate an ``effective'' energy barrier of sulfur jumps from the slopes in Figure~\ref{fig:jumps_arrhenius}.
For two vacancies in the structure (blue symbols), the   ``effective'' energy barrier of 0.81~eV agrees very well with the ``static'' value from \ac{NEB} calculations (0.83~eV).
In contrast for three vacancies (green symbol) the energy barrier of 0.63~eV derived from MD simulations by the  Arrhenius law fit differs by 0.22~eV from the value of 0.85~eV obtained from \ac{NEB} calculations.
This discrepancy highlights the significance of dynamical effects which are not captured in static \ac{NEB} calculations, such as vibrational contributions from neighboring atoms, and cooperative mechanisms involving simultaneous jumps.
As a result, these effects would not appear in typical \ac{kMC} simulations based on \ac{NEB}-calculated energy barriers but are taken into account in \ac{MD} simulations.

Having focused on temporal and averaged aspects of the sulfur-vacancy dynamics, we analyze next the site-dependent jump probabilities near small vacancy clusters.
This analysis is based on 10~ns long MACE \ac{MD} simulations with different initial vacancy configurations, see Figure~\ref{fig:jumpvis_mace}.
For all cases and consistent with earlier findings of \myAuthor{Wang} \textit{et al.}~\cite{Wang2022}, the sulfur-atom mobility is predominantly governed by the jump mechanism discussed in detail in  the context of Figure~\ref{fig:neb}.
As discussed above, this jump mechanism requires the presence of at least two adjacent sulfur vacancies. Furthermore, the jumping sulfur atom must be a first neighbor to both vacancies, and the interstitial site along the jump path must not be blocked by a molybdenum atom in the underlying atomic layer.  
Henceforth, we shall refer to these conditions as "hopping rules" or "vacancy-assisted hopping."

In the case of small vacancy clusters, the hopping rules constrain sulfur-atom mobility to a localized region surrounding the vacancy cluster.
These turn out to be equilateral triangles, as illustrated in Figure~\ref{fig:jumpvis_mace}~(a) and (b) for two or three vacancy clusters in the structure. 
According to the hopping rules and confirmed by the \ac{MD} simulations, all sulfur-atom jumps occur in a triangular pattern involving 3 or 6 sulfur lattice sites in total.
This explains the difference in \ac{MSD} between these two cases that has been mentioned earlier in the context of Figure~\ref{fig:msd}.
Notably, on the longer timescales accessible with the \ac{MLIP}, all jumps within the vacancy triangle occur with comparable frequency, as indicated by the gray connecting lines in Figure~\ref{fig:jumpvis_mace}.
Sulfur atoms outside of the outer triangles are not mobile because they either neighbor only one potentially vacant lattice site and/or a molybdenum atom blocks the path to adjacent vacancies.

The hopping rules observed here closely resemble so-called kinetically constrained models, which have long been used to investigate the dynamics of glassy systems and continue to be highly relevant in recent years~\cite{Ritort2003, Causer2020}.
In particular, the model of one-vacancy-assisted hopping on a triangular lattice of \myAuthor{Jäckle and Krönig}~\cite{Jckle1994, Kronig1994} differs from our hopping rules for MoS\textsubscript{2} only by the additional constraints imposed by the underlying molybdenum layer.
\myAuthor{Jäckle and Krönig} show that for their model of two-vacancy-assisted hopping, certain vacancy configurations can and will grow with probability 100\% across the whole system. They argue that for an infinitely large system this can lead to regular diffusion after a possibly very long cross-over time of subdiffusive propagation~\cite{Jckle1994, Kronig1994}.
Similarly, we will show that sulfur vacancy clusters in Mo\textsubscript{2} can capture vacancies just outside their triangular mobile region, creating a larger triangular mobile region and thereby facilitating sulfur vacancy mobility potentially across the whole system.
As illustration of this mechanism,  Figure~\ref{fig:jumpvis_mace}~(c) shows the jump probabilities resulting from a 10~ns long \ac{MD} simulation starting with three adjacent vacancies in a row (as in panel~(b)) and an additional vacancy, to be incorporated later. 
During the simulation, the three vacancies move within the 6-site mobile region of the cluster (c.f. panel (b)) and as soon as at least one of them is adjacent to the vacant site outside that region, a sulfur atom from outside the mobile region can jump.
This leads to cluster agglomeration, resulting in an expanded triangular mobile region of the cluster that now contains a total of $N_\mathrm{vac}\cdot(N_\mathrm{vac}+1)/2 = 10$ 
sites.
For such an agglomeration to occur, the vacancy to be captured must be adjacent to the mobile region of an existing cluster.
The location of this additional vacancy determines the direction in which the mobile region of the cluster is expanded.

To demonstrate how this mechanism can lead to the formation of larger vacancy clusters at comparatively low vacancy densities, we performed a \ac{MD} simulation for the vacancy arrangement shown in Figure~\ref{fig:jumpvis_mace}~(d).
Initially, only vacancies 1 and 2 are nearest neighbors, allowing for the sulfur jump to occur with a low energy barrier only within the dashed, red triangle.
Then, the next nearest neighboring vacancy 3 is captured similar to the process shown in Figure~\ref{fig:jumpvis_mace}~(c), resulting in a cluster of edge length 3, which is now adjacent to vacancy 4.
In this way, all vacancies are captured one by one, spanning the mobile region of the cluster through the periodic boundary of the supercell back to the initial vacancy 1, as indicated by the site-dependent jump rates shown in Figure~\ref{fig:jumpvis_mace}.

This vacancy-capture mechanism has not been observed in a previous \ac{MD} study of sulfur-vacancy dynamics by \myAuthor{Gao} \textit{et al.}~\cite{Gao2021}, most likely due to the inaccurate energy barriers of their classical reactive force field, as mentioned earlier in Subsection~\ref{subsec-otherhopps}.
Specifically, the ReaxFF force field overestimates the barrier for a sulfur jump into the edge of a three-vacancy line and underestimates the barrier for jumps into the center of the line---both by several tenths of an eV.
In contrast, the MACE \ac{MLIP} model developed here predicts these energy barriers with good accuracy compared to \ac{DFT}, enabling it to reliably reproduce the collective vacancy dynamics over long timescales and thus reveal the mechanisms behind the experimental observation of extended sulfur vacancy clusters~\cite{Komsa2013,Chen2018}, as will be discussed in detail below.

The panels (e) and (f) of Figure~\ref{fig:jumpvis_mace} are added to highlight the need for accurate \ac{MD} simulations capable of reaching timescales far beyond those accessible by \ac{AIMD}.
These panels present the site-resolved jump rates extracted from only the first 100~ps of the same \ac{MLIP}-based \ac{MD} simulations shown in panels (c) and (d), respectively.
The 100~ps timescale is representative of what can typically be achieved with \ac{AIMD} simulations.
For our typical MoS\textsubscript{2} supercells and using our CP2K setup (see Section~\ref{sub:methods_aimd}), a single iteration in the \ac{DFT} \ac{SCF} loop takes approximately 2.5~s on 24 CPU cores, with each \ac{MD} step requiring around seven \ac{SCF} iterations to converge the forces.
Consequently, simulating a 100~ps trajectory with a 0.5~fs timestep would require roughly 40~days computation time.
As illustrated in Figure~\ref{fig:jumpvis_mace}~(e) and (f), only a very small number of atomic jumps are observed within this timescale, predominantly near regions with a high initial vacancy concentration.
In contrast, about 1~ns of \ac{MD} can be computed within a day with both \ac{MLIP} architectures employed in this study. 

\subsection{Discussion of sulfur vacancy aggregation}

The concept of vacancy-assisted hopping expressed by the simple hopping rules explain the formation of regions with mobile sulfur atoms and sulfur vacancies. 
However, it is not sufficient to explain experimental facts such as the frequent observation of extended straight vacancy lines using \ac{TEM}~\cite{Komsa2013, Chen2018}, because it includes neither energy differences between different vacancy arrangements nor arrangement-dependent differences of the transition-barriers heights. 
Both these aspects are, of course, accounted for by the MLIP MD simulations and are discussed in detail in the following.

\begin{figure}[htpb]
    \centering
    \includegraphics[width=.48\textwidth]{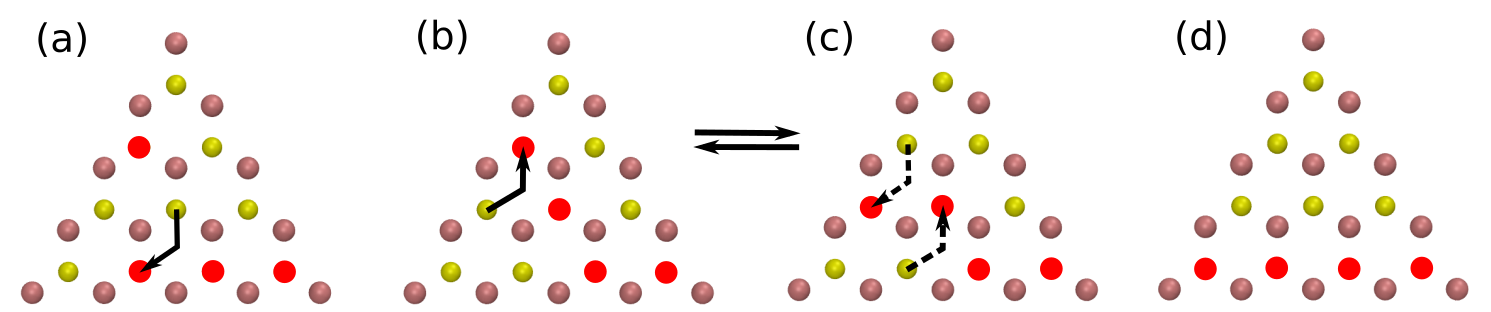}
    \caption{
    Illustration of a vacancy cluster of size 3 capturing a fourth vacancy, as observed in the \ac{MD} simulation underlying Figure~\ref{fig:jumpvis_mace}~(c).
    Following the capture after about 15~ps, the resulting four-vacancy cluster transitioned repeatedly between the configurations shown in panels (b) and (c) or similar ones, persisting in this dynamic state for a relatively long duration ($\sim$~3~ns) before eventually forming a straight line defect (d).
    The jumping sulfur atoms (yellow) are indicated by arrows.
    }
    \label{fig:vaccapture}
\end{figure}

The observed sequence of atomic jumps leading to the incorporation of additional vacancies into an existing cluster is depicted in \textbf{Figure~\ref{fig:vaccapture}}, based on the same \ac{MD} simulation as Figure~\ref{fig:jumpvis_mace}~(c).
In this case, the fourth vacancy, initially located outside the mobile region of the original three-vacancy cluster, is captured relatively fast, i.e. within approximately 15~ps.
This results in the formation of a vacancy line with a 60$^\circ$ kink (Figure~\ref{fig:vaccapture}~(b)).
The resulting configuration persists in a dynamic equilibrium with other similar structures, most notably the one shown in Figure~\ref{fig:vaccapture}~(c).
Both jumps indicated in Figure~\ref{fig:vaccapture}~(c) lead to configurations that are symmetry-equivalent to the one shown in panel (b).
Notably, once the buckled four-vacancy line forms, no dissociation into smaller clusters is observed.
Eventually, after around 3~ns, the cluster forms a straight line of four vacancies located along the edge of the triangular mobile region visible in Figure~\ref{fig:jumpvis_mace}~(c).
This configuration remains comparably stable thereafter, with only occasional, brief jumps of vacancies back out of the line, followed by their prompt return.

\begin{figure}[htpb]
   \centering
    \includegraphics[width=.45\textwidth]{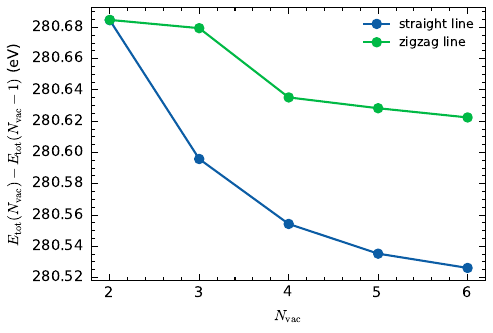}
    \caption{
    Difference in \ac{DFT} total potential energy between linear and zigzag arrangements of $N_\mathrm{vac}$ and $N_\mathrm{vac}\!-\!1$ sulfur vacancies.
    As detailed in Section~\ref{sub:formation_energy}, the values presented here correspond to the difference in defect formation energies between lines of length $N_\mathrm{vac}$ and $N_\mathrm{vac}\!-\!1$, offset by the sulfur chemical potential.
    }
    \label{fig:formation_energy}
\end{figure}

The tendency of vacancy clusters to align in straight lines along the edge of their mobile region is consistently observed across our \ac{MD} simulations.
To explore the driving force behind this behavior, we analyze the total DFT energy difference $E_\mathrm{tot}(N_\mathrm{vac})-E_\mathrm{tot}(N_\mathrm{vac}\!-1)$between vacancy lines of lengths $N_\mathrm{vac}$ and $N_\mathrm{vac}\!-\!1$, which differs only by the sulfur chemical potential $\mu_S$ from the differences in the formation energy; see Section~\ref{sub:formation_energy}.
The energy difference can be interpreted as the binding energy of an additional vacancy to a preexisting line of $N_\mathrm{vac}\!-\!1$ vacancies.
\textbf{Figure~\ref{fig:formation_energy}} shows that  the vacancy binding energy decreases monotonically with increasing vacancy line length, almost certainly approaching a finite value in the large-$N_\mathrm{vac}$ limit.
Therefore, the formation energy is a concave curve and consequently longer vacancy lines are energetically more favorable than multiple shorter lines with the same total number of vacancies. 
This general result is in full agreement with the conclusions of \myAuthor{Le} \textit{et al.}~\cite{Le2014} and \myAuthor{Komsa} \textit{et al.}~\cite{Komsa2013}, who compared the formation energies of vacancy lines of varying lengths at a specific value of the sulfur chemical potential.

Figure~\ref{fig:formation_energy} also includes these energy differences for zigzag configurations as a function of the number of vacancies $N_\mathrm{vac}$ (green symbols).
In the zigzag configuration, the angle between any three adjacent vacancies is 120$^\circ$, as shown for $N_\mathrm{vac} = 4$ in  Figure~\ref{fig:vaccapture}~(c).
The zigzag energy differences are consistently larger than their straight-line counterparts (blue symbols).
Taken together, these two observations explain the tendency of vacancy clusters to organize into long lines along the edge of their mobile region as observed in our \ac{MD} simulations.
This formation of extended vacancy lines has also been observed experimentally using \ac{TEM}~\cite{Komsa2013, Chen2018} and is believed to play a crucial role in the functional behavior of memristive elements \cite{Jadwiszczak2019}.

We need to address three more aspects of vacancy dynamics, before we compare to what is experimentally known. 
The first aspect concerns cluster growth. We saw that clusters can grow by capturing additional vacancies located adjacent to their mobile region (cf. Figure~\ref{fig:vaccapture}). For the following argument, which is adapted from Ref.~\cite{Jckle1994}, we ignore any energetic aspects and consider only the hopping rules and argue that for any finite vacancy concentration $c$ and infinite systems there will be with probability 100\% a mobile region that can grow across the whole system. 
The chances that for triangular mobile region with edge length $N_{\textrm{vac}}$ at least one of its $3(N_{\textrm{vac}}+1)$ neigboring sites is vacant, is
$1-(1-c)^{3(N_{\textrm{vac}}+1)}$. 
This factor approaches $1$ exponentially, as soon as $N_{\textrm{vac}}\gg 1/(3c)$.
Completely analogously to the growth of rain drops in oversaturated air, it is difficult for a condensation nucleus to reach a critical size; but once the critical size has been reached, further growth is almost certain.
For example, the mobile region of a vacancy line with length $N_\mathrm{vac} = 4$ has 15 adjacent sulfur lattice sites.
Experimentally, vacancy densities of $\rho_\mathrm{vac} = N_\mathrm{vac} / N_\mathrm{S~sites} = 0.1$ can be achieved by irradiation while retaining structural properties of MoS\textsubscript{2} \cite{Ma2013, Bertolazzi2017}.
Under these conditions, the probability that at least one of the 15 neighboring sites is vacant---thus allowing the cluster to eventually capture an additional vacancy---is $1 - 0.9^{15} \approx 80 \%$.

\begin{figure}[htpb]
    \centering
    \includegraphics[width=.48\textwidth]{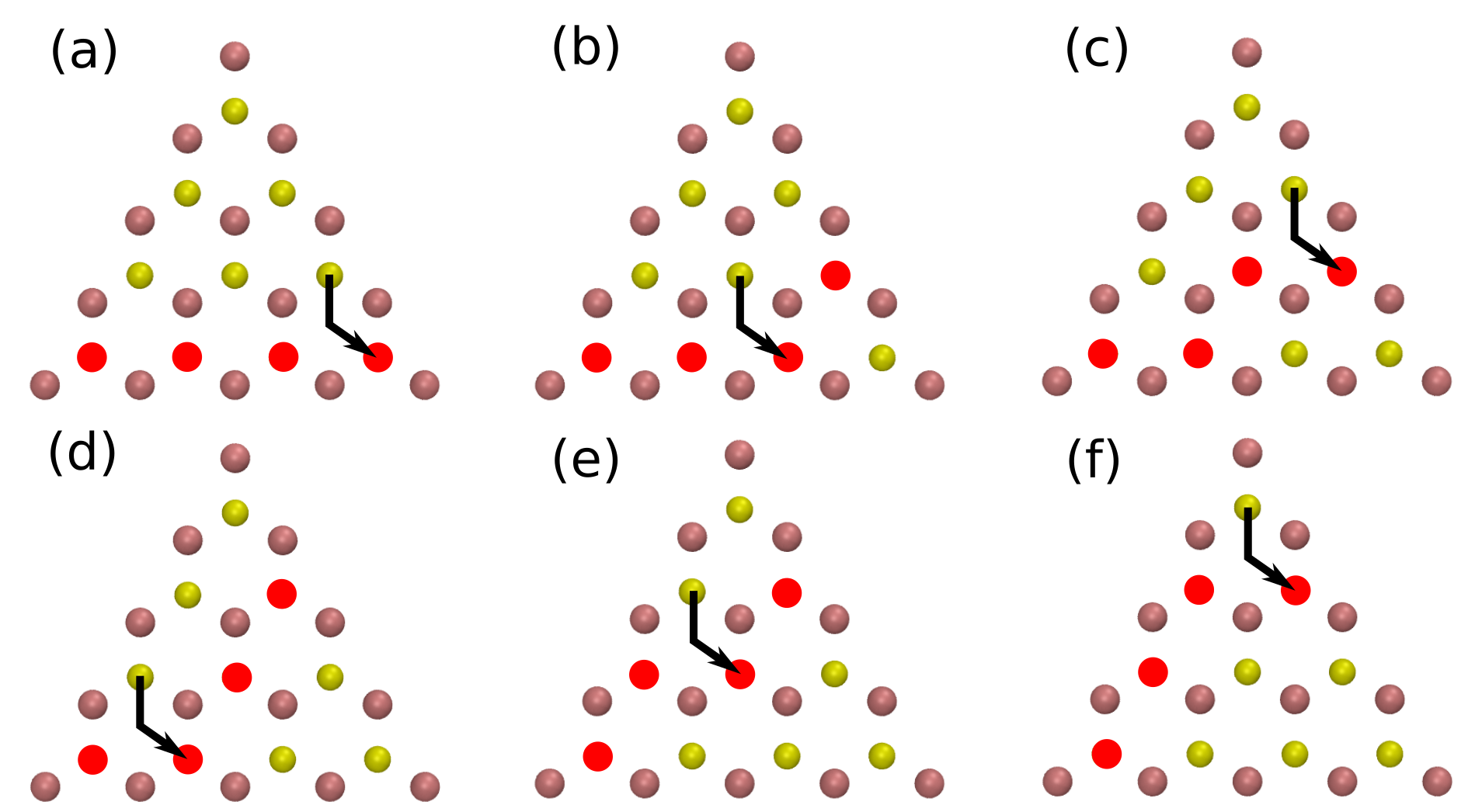}
    \caption{
    Illustration of the most direct series of low-energy jumps leading to the flip of a row of four vacancies (red) within its mobile region (cf. Figure~\ref{fig:jumpvis_mace}~(c)).
    }
    \label{fig:rowflip}
\end{figure}

The second aspect concerns the mobility of line defects:
We saw that most likely a straight line is the energetically most advantageous configuration of a cluster involving a given number of vacancies (cf. Figure~\ref{fig:formation_energy}).
However, at finite temperature in \ac{MD} simulations or sample growth, entropy will disfavor this configuration.
Indeed, in MD simulations, we typically observe a dynamic equilibrium in which the vacancy cluster is located near the edge of the mobile region. There, buckled configurations---similar to those shown in  Figure~\ref{fig:vaccapture}~(b) and (c)---continue to form before transitioning back to a straight line.
Over longer timescales, on the order of several nanoseconds for small $N_{\textrm{vac}}$, the vacancy line can even relocate to a different edge of the mobile region via a sequence of many individual jumps.
The most straightforward such sequence obeying the hopping rules is illustrated in \textbf{Figure~\ref{fig:rowflip}}.
Noticeably, during some \ac{MD} runs, we observe successive jumps---following for example the mechanism illustrated in Figure~\ref{fig:rowflip}~(b) and (c)---occurring within a short time span ($\sim 1$~ps).
In these cases, a sulfur atom initially blocking the jump of a second sulfur atom relocates just prior to or concurrently with the second jump, enabling both jumps to proceed in a correlated manner.
This observation highlights the value of \ac{MLIP} \ac{MD} simulations over extended timescales, as other approaches---such as \ac{kMC} models---require prior knowledge of such correlated vacancy dynamics in order to accurately capture cooperative vacancy jumps.

The third aspect concerns the question whether the energetically advantageous line defects break into pieces at finite temperature in order to increase entropy. 
This is a purely academic question for very long chains, because their formation along the lines described above takes too long, both in the computer and in real life.  
In our \ac{MLIP} \ac{MD} simulations with $N_{\textrm{vac}}\lesssim 6$, we do not observe lines splitting apart after their initial formation.
Occasionally, individual sulfur atoms jump into vacancies near the center of a line, transiently dividing it into two segments, but they return to their original sites before further jumps can stabilize the new configuration and lead to a permanent line reconfiguration.
To investigate the energetic stability of vacancy lines, we performed further \ac{DFT} calculations comparing atomic configurations to assess the driving forces opposing line fragmentation beyond the data already shown in Figure~\ref{fig:formation_energy}.
For straight line defects with $N_{\textrm{vac}}=5$ ($N_{\textrm{vac}}=10$), the energetic costs for a sulfur atom to jump into the center of the line are about 0.40~eV (0.59~eV) wheres jumps into one of the terminal vacancies cost only about 0.15~eV (0.20~eV).  
For $N_{\textrm{vac}} = 10$, a partial 60$^\circ$ flip of half the vacancy line (Figure~\ref{fig:rowflip})---dividing it into two segments of five vacancies---yields a configuration 0.68~eV higher in energy than the single straight line.
In summary, in the regime $N_{\textrm{vac}}\lesssim 10$, static \ac{DFT} calculations confirm the presence of rather strong energetic driving forces that counteract the entropically driven disintegration of vacancy lines, explaining and confirming our observation of stable vacancy lines in nanosecond-scale \ac{MD} simulations.

The non-negligible variation in formation energies and energy barriers across different vacancy configurations---as discussed here and illustrated in Figure~\ref{fig:formation_energy}---underscores a key advantage of the \acp{MLIP} developed in this work over previously used \ac{kMC} models for studying sulfur dynamics in MoS\textsubscript{2}.

The formation of vacancy clusters has been analyzed experimentally at the atomic scale using \ac{TEM} imaging.
\myAuthor{Komsa} \textit{et al.}~\cite{Komsa2013} observed the agglomeration of irradiation-induced vacancies into a large number of line defects at room temperature.
According to our \ac{MD} simulations, this can be attributed to small vacancy clusters incorporating a nearby vacancy through a short sequence of low-energy jumps as illustrated in Figure~\ref{fig:vaccapture}.
But with increasing cluster size, this takes more and more time, as it requires a specific sequence of individual jumps to occur in the right order for the vacancies to hop to another edge of their mobile region to capture further vacancies. 
As a result, many moderately sized vacancy clusters form at room temperature, generally aligning along straight lines~\cite{Komsa2013}.
At elevated temperatures and, thus, larger hopping rates, \ac{TEM} measurements reveal the emergence of substantially longer vacancy lines, extending over tens of nanometers~\cite{Chen2018}.
Insights from our \ac{MLIP} \ac{MD} simulations strongly suggest that higher temperatures enhance the mobility of moderately long lines by allowing them to overcome the energetic barriers associated with intermediate configurations during the line-flip process (see Figure~\ref{fig:rowflip}).
Consequently, as clusters grow, longer lines can still incorporate additional vacancies adjacent to their triangular mobile regions, as illustrated for a small vacancy cluster in Figure~\ref{fig:vaccapture}.
These temperature-dependent constrained defect dynamics may lead to glass-like behavior with history-dependent properties, as seen in the kinetically constrained models briefly discussed above~\cite{Ritort2003, Causer2020,Jckle1994, Kronig1994}.

\section{Conclusion}
Large sulfur vacancy clusters in MoS\textsubscript{2} form and undergo dynamic changes even at elevated temperatures on timescales that are inaccessible to AIMD calculations. 
The underlying processes depend on details of the cluster configuration in a way that is too complex for \ac{kMC} simulations.
This leaves MD based on \acp{MLIP} as, arguably, only alternative.
However, such calculations are still challenging challenge because the transition states and energy barrier heights between (meta-) stable positions must be described quantitatively.
We find that \acp{MLIP} can accurately describe the potential energy surface near stable, minimum-energy configurations of MoS\textsubscript{2} with only minimal fine-tuning, see Figure~\ref{fig:neb}. 
In contrast, a careful construction and selection of appropriate training datasets is needed for a qualitatively and quantitatively correct description of the transition state.

To investigate this issue, we compared two \ac{MLIP} architectures: \ac{GAP} and the equivariant \ac{GNN} MACE~\cite{Bartok2010, Jinnouchi2019, Jinnouchi2019a, Jinnouchi2020, Batatia2022}.
We assessed different dataset generation strategies by evaluating potential energy profiles for the lowest-energy sulfur atom jump, as predicted by \acp{MLIP} trained on different datasets, and compared them to reference \ac{DFT} calculations.

The GAP framework inherently supports on-the-fly training strategies, allowing it to dynamically filter atomic structures for inclusion in the training set.
This leads to efficient training requiring relatively few \ac{DFT} calculations and no ``manual'' selection of representative structures for the training set to achieve accurate results.
In contrast, the accuracy of MACE was found to be more sensitive to the size and representativeness of the training dataset.
We successfully trained a MACE \ac{MLIP} on a large dataset sampled from \ac{AIMD} simulations.
This approach yielded highly accurate predictions for both the transition-state energies for the important 'vacancy-assisted hopping' and the \ac{RDF}.
Finally, both approaches reliably reproduce the underlying potential energy surfaces that govern vacancy migration. This suggests that usability---understood as simplicity of application and efficiency in data requirements---may outweigh minor differences in accuracy when it comes to determining which MLIP approach will be most widely adopted in the future.


Importantly, training accurate \ac{MLIP} models allowed us to perform \ac{MD} simulations far beyond the typical time scales that have previously been accessible at \textit{ab initio} accuracy and thus enabled us to gain new insights into the local dynamics of sulfur vacancies in MoS\textsubscript{2}.
These dynamics are driven by a specific low-energy sulfur jump mechanism ('vacancy-assisted hopping') occurring adjacent to at least two vacancies~\cite{Komsa2013, Le2014, Wang2022}.
By analyzing \acp{MSD}, jump rates, and site-dependent jump probabilities from nanosecond-long \ac{MLIP}–\ac{MD} simulations, we observed new mechanisms behind the collective mobility of sulfur vacancies at the atomic level.
These atomistic processes are inaccessible to static \ac{DFT} calculations or \ac{kMC} models, highlighting the added value of long-timescale, high-accuracy simulations.

At low vacancy densities, only small clusters form.
The mobility of such clusters is confined to a triangular region, with an edge length approximately equal to or less than the number of vacancies within the cluster due to the specific sulfur jump governing the vacancy mobility~\cite{Komsa2013, Le2014, Wang2022}.
By computing ``effective'' energy barriers via the Arrhenius relation, we find that the mobility within these clusters is strongly size-dependent.
This analysis of \ac{MD} simulations at different temperatures also highlights the importance of dynamical effects---such as vibrational contributions from neighboring atoms, and cooperative mechanisms involving simultaneous jumps---which could not be captured by the combination of static \ac{NEB} calculations and \ac{kMC} simulations previously used to study sulfur dynamics in MoS\textsubscript{2}~\cite{Wang2024,Wang2022,Wang2024a}.

Furthermore, we demonstrate that small clusters can capture vacancies located adjacent to their restricted area of motion, thereby expanding their range.
This mechanism could in principle lead to large and even infinite, mobile vacancy clusters at experimentally relevant vacancy densities, but would need very long to do so.

MD simulations and static DFT calculation indicate that within these mobile regions, vacancies preferentially form straight lines in order to minimize their energy.  
Taken together, these findings explain the experimental observation of a large number of moderately long vacancy lines at room temperature, and fewer---but significantly longer---vacancy lines, spanning several tens of nanometers, at elevated temperatures~\cite{Komsa2013, Chen2018}.
The mobility of small vacancy clusters observed here suggests that vacancies could preferentially agglomerate at grain boundaries, where a higher defect density would permit longer‐range diffusion. This highlights the need for further theoretical modeling of vacancy dynamics in polycrystalline MoS\textsubscript{2} systems.

We expect future research to explore the interwoven aspects of concentration dependence, history-dependent glassy behavior, and memristivity on large scales in more detail.
These all originate from the non-linearity intrinsic to any vacancy-assisted (vacancy) propagation.
Likely, effective continuum models will be developed for that purpose based on concentration- and history-dependent mobilities derived from \ac{MLIP} \ac{MD}.


\section{Methods}
\subsection{\textit{Ab initio} energies and molecular dynamics}
\label{sub:methods_aimd}

The \ac{AIMD} data used to fine-tune MACE models was calculated with the CP2K software package~\cite{Khne2020} using the Quickstep algorithm \cite{VandeVondele2005} and efficient orbital transformation \cite{VandeVondele2003} with  Gaussian basis sets~\cite{VandeVondele2007} and
pseudopotentials~\cite{Hartwigsen1998} at a kinetic energy cutoff of 500~eV.
To generate training data for \ac{GAP} \acp{MLIP} and calculate potential energy curves, we also performed \ac{DFT} calculations using the Vienna \textit{Ab initio} Simulation Package (VASP~6.4)~\cite{Kresse1996, Hafner2008} within the projector augmented-wave method \cite{Blchl1994} with a kinetic energy cutoff of 430~eV.
The PBE exchange correlation functional~\cite{Perdew1996} was used throughout this study. 
A single $\mathbf{k}$ vector at the $\Gamma$ point is expected to suffice for all integrations in the reciprocal $\mathbf{k}$ space because the studied MoS\textsubscript{2} supercells are large in real space. 

The supercells contain typically 108 atoms in a  monolayer of hexagonal 2H-MoS\textsubscript{2} parallel to the x-y plane.
The z dimension of the supercell was chosen to be 40~\AA{} 
in order to minimize the interaction between replica of the monolayer originating from periodic boundary conditions.
A larger supercell with 432 atoms is used in order to reduce the in-plane interaction between periodic images of the 4 vacancy clusters shown in Figure~\ref{fig:jumpvis_mace}~(c) and (e).

All calculations presented in this work were performed without charge compensation, i.e., as 'undoped' systems, because we found the neutral sulfur vacancy to be energetically favorable for electron chemical potentials near the center of the band gap when comparing supercells with sulfur vacancies for different numbers of electrons and compensating background charges.
This is in agreement with previous \ac{DFT} results~\cite{Komsa2015}.

The \Ac{MD} simulations were performed within the NVT ensemble and with the Nosé-Hoover-Thermostat~\cite{NOS2002, Nos1984, Martyna1992} to keep the temperature constant at, in most cases,  1000~K.
All simulations were done within the Born-Oppenheimer approximation.
We chose a \ac{MD} time step of 0.5~fs for production runs and 1~fs for simulations used to generate training data for \acp{MLIP}.

\subsection{Training of \ac{MLIP} models}
\label{sub:methods_mlff}

\subsubsection{MACE}
As a starting point for training MACE models~\cite{Batatia2022}, we use the MACE MP-0 foundation model~\cite{Batatia2023} which has been pre-trained on \ac{DFT} data of about 150,000 crystals extracted from the materials project database~\cite{Jain2013}.
To generate an accurate MACE model for MoS\textsubscript{2}, we employed two different approaches for generating training data used to fine-tune the foundation model.
(\textit{i}) We performed a 10~ns long \ac{MD} simulation using the MACE foundation model with two adjacent sulfur vacancies in the initial structure.
Then, we calculated \ac{DFT} energies and forces on a series of equidistantly selected snapshots.
This test dataset for the foundation model was then used to fine-tune the MACE model.
(\textit{ii}) Alternatively, we extracted \ac{DFT} energies and forces directly from a 90~ps long \ac{AIMD} trajectory with the same initial structure.

For both approaches, we compared \acp{MLIP} fine-tuned to different numbers of training structures which have been sampled equidistantly from the respective \ac{MD} simulations.
As discussed in the Results Section, we find the latter approach to yield a more accurate \ac{MLIP} and thus used it in our production runs.
Additionally, we find that increasing the fine-tuning set size to a large number of 50000 structures yields noticeably better results.
In the future, this number could be reduced by sampling the training data for rare events, possibly without affecting the accuracy of the \ac{MLIP}.
During each training process, 10~\% of the training data was reserved for validation.
Since these \ac{MLIP} models are designed for use in \ac{MD} simulations, our loss function contained a weight of the force error that is 100 times that of the energy error.

\subsubsection{\Ac{GAP}}
Each of the \acp{GAP}~\cite{Bartok2010, Jinnouchi2019, Jinnouchi2019a, Jinnouchi2020} was trained in a on-the-fly learning MD run with a simulation length of 1~ns and the \ac{MD} parameters described in Section~\ref{sub:methods_aimd}.
The \acp{GAP} were refitted to new \ac{DFT} data on the current structure of the \ac{MD} simulation whenever the Bayesian error estimate for the force on at least one atom exceeded a given threshold.
This threshold value was initialized at 2~meV~A$^{-1}$ and then dynamically updated during the on-the-fly training.
For these update steps the \ac{GAP} to \ac{DFT} error at the previous 10 force field refits were taken into account.
In this way, rare events are automatically overrepresented in the training dataset, thus requiring fewer structures in the training dataset compared to our most accurate MACE model.
Typical \ac{GAP} runs triggered roughly 600 on-the-fly  \ac{DFT} calculation with subsequent adjustment of the \ac{MLIP}.

Again, we compared two different approaches for selecting training structures:
(\textit{i}) Four separate \acp{MLIP} were trained for four distinct defective MoS\textsubscript{2} structures: those containing one, two, and three sulfur vacancies in a row, as well as a MoS\textsubscript{3} vacancy, i.e.\ the defect resulting from the removal of one molybdenum atom along with its three adjacent sulfur atoms in one of the sulfur layers.
(\textit{ii}) Alternatively, a single MLIP was obtained by subsequent GAP calculations for these four defect structures, where the training run for any one of these defect structures was started with the \ac{MLIP} resulting from training on the previous defect structure.
Thus, the resulting  \ac{MLIP} had  multiple, quite distinct defect structures in its training database.

\subsubsection{Test errors}
\label{sub:methods_testerror}


For both \ac{MLIP} architectures, the test set was generated after the training process by sampling structures from an \ac{MD} simulation performed using the trained model.
DFT forces and energies were then computed for 150 equidistantly sampled snapshots from this trajectory.
The resulting force and energy test errors for selected \acp{MLIP} models are presented in \textbf{Table~\ref{tab:testtrainingerrors}} and discussed in Section~\ref{sub:results_mliptraining}.
\begin{table}[htpb]
 \centering
 \caption{
Test errors (root-mean-square error: RMSE) of selected \acp{MLIP}. For the MACE architecture, results are shown for two models. One trained on 50,000 snapshots from an \ac{AIMD} simulation and the other on 2,000 snapshots from an \ac{MD} simulation using the MACE MP-0 foundation model --- both featuring two adjacent sulfur vacancies in the initial structure. The \ac{GAP}-based \acp{MLIP} were trained on-the-fly, either using four different defect structures (see Section~\ref{sub:methods_mlff}) or a single structure with two ($S_2$) or three ($S_3$) adjacent sulfur vacancies. Test errors were computed from 150 snapshots sampled from 10~ns \ac{MLIP} \ac{MD} simulations, benchmarked against \ac{DFT} calculations. All \ac{MLIP} simulations used the same initial structures as those employed in the training or fine-tuning of the respective models. For the \ac{GAP} \ac{MLIP} trained on multiple defect structures, application tests were conducted using a structure containing three adjacent sulfur vacancies. 
 }
  \begin{tabular}{|@{}l|c|c@{}|}
    \hline ~Test error ~
    & energy  & force~~  \\
    ~(RMSE) &  meV/atom & meV/\AA~ \\
    \hline
    ~MACE 
    (test set 2000 structures) & 10 & 112 \\
    ~MACE 
    (AIMD 50000 structures) & 4.8 &  42.0 \\
    ~GAP 
    (multiple defect structures) &  26  &  142 \\
    ~GAP 
    ($S_2$ vacancy) & 29 & 73.8 \\
   ~GAP 
    ($S_3$ vacancy) & 23 & 85.9\\
    \hline
  \end{tabular}
  \label{tab:testtrainingerrors}
\end{table}

\subsection{Potential energy curves and activation energies}
\label{sub:methods_neb}
Figure~\ref{fig:neb} shows activation barriers and potential energy curves as function of the so-called configuration coordinate for a single sulfur atom moving from its initial position $\mathbf{r_0}$ into an interstitial site at $\mathbf{r_1}$ surrounded by three sulfur vacancies, as illustrated in the inset of Figure~\ref{fig:neb} (b). 
Due to the symmetric situation, the path of the sulfur atom is found to be very close to a straight line and, thus, well described by 
\begin{equation}
    \lambda = \frac{\left( \mathbf{r} - \mathbf{r_0} \right) \cdot \left( \mathbf{r_1} - \mathbf{r_0} \right)}{\left( \mathbf{r_1} - \mathbf{r_0} \right) \cdot \left( \mathbf{r_1} - \mathbf{r_0} \right)}
    \label{eq:confco}
\end{equation}
as configuration coordinate corresponding to the moving sulfur's position $\mathbf{r}$.
To find the minimal energy path between the start and end point of the transtition, the \ac{NEB} method was applied~\cite{Jonsson1998}.
To ensure comparability between the curves, $\mathbf{r_0}$ and $\mathbf{r_1}$ were always taken from the \ac{DFT} geometry optimizations using VASP, although the initial and final configurations of the \ac{NEB} were independently optimized for each method (CP2K, MACE, or GAP).  

\subsection{Radial distribution functions}
\label{sub:methods_rdf}
The S-S \acp{RDF} shown in Figure~\ref{fig:rdf} were calculated according to

\begin{eqnarray}
  g_{S-S} (r) &=& \frac{1}{4\pi r^2 \rho} \frac{1}{N_S} \sum_{i,j=1}^{N_S}
  \big<\, \delta ( \rvert \mathbf{r}_i - \mathbf{r}_j \lvert - r)  \,\big>_t \\
    &=& \frac{V}{N_S^2}
        \frac{\big<\,n_{S-S}(r)\,\big>_t}
        {4 \pi r^2 \Delta r}
\end{eqnarray}
Here, $N_S$ and $V$ denote the number of sulfur atoms in the supercell and its volume, respectively.
where $\rho=N_S/V$, $N_S$ and $V$ denote the sulfur density, the number of sulfur atoms in the supercell, and its volume, respectively.
The number of sulfur pairs with a given distance $n_{S-S}(r)$ was determined by constructing a histogram with a bin width of $\Delta r = 0.016$~\AA.
The time average was calculated over about 10,000 time steps of each \ac{MD} simulation shown in Figure~\ref{fig:rdf}.
The first picosecond of the simulation was left out.

\subsection{Mean squared displacements}
\label{sub:methods_msd}
The starting point of the MSD calculation is the typical definition
\begin{equation}
 \label{eq:msd_normal}
    \mathrm{MSD}(\tau) = \left<\, \left<
    \lvert \mathbf{r}_i^{(S)}(t_0 + \tau) - \mathbf{r}_i^{(S)}(t_0) \rvert^2
    \right>_i \right>_{t_0}
\end{equation}
Here, $\mathbf{r}_i^{(S)}(t)$ denotes the position of the i-th sulfur atom at time $t$.
The two averages are over all sulfur atoms in the system and over all time intervals with length $\tau$ starting from $t_0 \in [0,t_\mathrm{end} - \tau]$.

As described in the Results Section, the motion of sulfur atoms not close to vacancies is restricted to the relatively small surrounding.
Thus, in our case, averaging over all sulfur atoms creates an artificial dependence of the \ac{MSD} on the number of sulfur atoms in the supercell.
To eliminate this dependence, we adjust the definition from Equation~(\ref{eq:msd_normal}) in two different ways:

The \acp{MSD} in Figure~\ref{fig:msd}~(a) are not averaged but summed over sulfur atoms. We still refer to it as a \ac{MSD} because the mean displacements over all time windows with a given lag time has been calculated.
Additionally, the offset due to the atoms' vibrations about their equilibrium positions is subtracted.
The contribution to the squared displacement due to these thermal vibrations was calculated as the \ac{MSD} of a defect-free supercell for long lag times.
This way, only mobile atoms contribute to the \ac{MSD} and the MACE foundation model is comparable to the other two \acp{MLIP} even though it predicts too many atoms to be mobile.

The \acp{MSD} in Figure~\ref{fig:msd}~(b) are not averaged over all sulfur atoms, but only over the mobile sulfur atoms.
Atoms are categorized as mobile, if their maximum MSD during the span of the calculation exceeds half of the distance between two adjacent sulfur sites.
Both the \acp{MLIP} shown here predict one atom to be mobile in a cluster with two vacancies and three atoms to be mobile in a cluster with three vacancies.

\subsection{Defect formation energy}
\label{sub:formation_energy}

The stability of vacancy lines of different lengths $N_\mathrm{vac}$ can be assessed using their formation energies $E_F$.
We define the energy for the formation of extended vacancy line defects by using the \ac{DFT} energy of a single sulfur vacancy as reference:
\begin{equation}
    \label{eq:formation_energy}
    E_F(N_\mathrm{vac}) = E_\mathrm{tot} (N_\mathrm{vac}) - E_\mathrm{tot}(1) + (N_\mathrm{vac} - 1) \mu_S 
\end{equation}
$E_\mathrm{tot} (N_\mathrm{vac})$ and $E_\mathrm{tot}(1)$ denote the total potential energies of a structure containing a vacancy line with length $N_\mathrm{vac}$ and of the reference structure containing just a single vacancy, respectively.
The sulfur chemical potential $\mu_S$ depends strongly on the chemical environment of the MoS\textsubscript{2} sheet.
In this case, in order to compare the stability of vacancy lines with varying length it is sufficient to analyze differences in the formation energy.
\begin{align}
    \label{eq:delta_formation_energy}
    \Delta E_F(N_\mathrm{vac}) &= E_F(N_\mathrm{vac}) - E_F(N_\mathrm{vac} - 1)\\
    &= E_\mathrm{tot} (N_\mathrm{vac}) - E_\mathrm{tot} (N_\mathrm{vac} - 1 ) + \mu_S\\
    &= \Delta E_\mathrm{tot} (N_\mathrm{vac}) + \mu_S
\end{align}
It is evident that both $\Delta E_F(N_\mathrm{vac})$ and $\Delta E_\mathrm{tot}(N_\mathrm{vac})$ have the same slope. This allows for a qualitative assessment of the curvature of $E_F(N_\mathrm{vac})$ without making any assumptions about the sulfur chemical potential, as discussed in the Results Section in the context of Figure~\ref{fig:formation_energy}.
These calculations were carried out using the CP2K software package as described in Section~\ref{sub:methods_aimd}.

\medskip

\medskip
\textbf{Data availability statement}\par
The data presented and discussed here, as well as the trained \ac{MLIP} models are available from the corresponding author upon request.
\medskip

\textbf{Funding statement} \par 
Support through the ``Ilmenau School of Green Electronics'' that was made possible by funding from the Carl-Zeiss-Stiftung is gratefully acknowledged (Projekt P2022-00-135).
\medskip

\textbf{Conflict of interest disclosure}\par
The authors declare no conflict of interest.
\medskip

\textbf{Acknowledgements} \par 
The authors thank Henning Schwanbeck and his colleagues at the University Computing Center of the TU Ilmenau for excellent working conditions and continued support.

\medskip
\bibliography{exportNoAbstract}

\end{document}